\documentclass[authoryear,preprint,review,12pt]{elsarticle}

\usepackage{amssymb}
\usepackage{amsthm}

\usepackage{lineno}

\usepackage[utf8]{inputenc}
\usepackage{csvsimple}
\usepackage{gensymb} 
\usepackage{amsmath}
\usepackage{float}
\usepackage{natbib}
\usepackage{color, colortbl}
\usepackage{geometry}
\usepackage{tabularx}
\usepackage{multirow} 
\usepackage{multicol} 

\definecolor{Gray}{gray}{0.9}

\graphicspath{{Figures/}}

\usepackage{hyperref}

\journal{Planetary and Space Science}

\begin{document}

\begin{frontmatter}{}

\title{Image processing for precise geometry determination}

\author[label1,label2]{I. Belgacem\corref{cor1}}
\ead{ines.belgacem@esa.int}

\author[label3]{G. Jonniaux}
\author[label1]{F. Schmidt}

\address[label1]{Université Paris-Saclay, CNRS, GEOPS, 91405, Orsay, France.}
\address[label2]{European Space Astronomy Centre, Urb. Villafranca del Castillo, E-28692 Villanueva de la Cañada, Madrid, Spain}
\address[label3]{Airbus Defence $\&$ Space, Toulouse, France.}

\cortext[cor1]{Corresponding author}

\begin{abstract}
Reliable spatial information can be difficult to obtain in planetary remote sensing applications because of errors present in the metadata of images taken with space probes. We have designed a pipeline to address this problem on disk-resolved images of Jupiter's moon Europa taken with New Horizons' LOng Range Reconnaissance Imager, Galileo's Solid State Imager and Voyager's Imaging Science Subsystem. We correct for errors in the spacecraft position, pointing and the target's attitude by comparing them to the same reference. We also address ways to correct for distortion prior to any metadata consideration. Finally, we propose a vectorized method to efficiently project images pixels onto an elliptic target and compute the coordinates and geometry of observation at each intercept point. 
\end{abstract}


\end{frontmatter}{}


\section{Introduction}
For a variety of applications in space exploration remote sensing, it is crucial to have accurate spatial representation of the data. To do so, the user needs precise information about the position and attitude of the spacecraft. The SPICE information system \citep{Acton_AncillaryDataServices_PaSS_1996} helps planetary scientists and engineers to plan and process remote sensing observations. Groups in the different missions create Camera-matrix Kernels (CK) and Spacecraft Kernels (SPK) for SPICE containing the pointing information of the instrument of and the position of the spacecraft. However, the data set to generate these kernels incorporates uncertainties and errors which can make it difficult to accurately project pixels into the 3D scene \citep{Jonniaux_RobustExtractionNavigation_GN&CS_2014, Sidiropoulos_SystematicSolutionMulti_IToGaRS_2018}. 

Efforts have been made to develop tools that correct and reconstruct the CK kernels using astrometry techniques to generate the so called C-smithed kernels  \citep{Cheng_PDS_NH_LORRI_ckinfo}. However, not all C-Kernels have undergone this correction and some of these corrections might be considered by-products and not necessarily be publicly available or easily accessible.

Some open-source tools are available such as ISIS3 \citep{Anderson_ModernizationIntegratedSoftware_LaPSC_2004, Edmundson_JIGSAWISIS3BUNDLE_IAoPRSaSIS_2012}, but it needs a manual selection of control points for optimal results. Additionally, we have verified after processing Voyager images with this tool that a lot of distortion was still present which is a major issue. The AutoCNet python library \citep{Laura_AutoCNetPythonlibrary_SX_2018} has been developed with the objective of a more automated tool based on dense feature extraction on high resolution images. The CAVIAR software package \citep{Cooper_CaviarSoftwarePackage_AA_2018} is using background star positions to refine spacecraft pointing information and is publicly available to correct CASSINI images metadata. There is also an on-going effort of the PDS Ring-Moon Systems Node to develop tools to easily interact with SPICE geometric metedata \citep{Showalter_DevelopmentsGeometricMetadata_IaDA_2018}.

We propose here solutions for correcting metadata in disk-resolved images at intermediate resolution, typically when the full planetary body is observed in a single scene. We hypothesize that the ephemeris of the planetary bodies involved are valid and therefore we correct for the pose of the cameras (position and attitude). We also propose an efficient way to project all pixels onto the target. To illustrate this work, we are using images of Jupiter's moon Europa taken with three different imagers - the New Horizons' LOng Range Reconnaissance Imager (LORRI) \citep{Cheng_LongRangeReconnaissance_SSR_2008} Galileo's Solid State Imager (SSI) \citep{Belton_GalileoSolidState_SSR_1992} and Voyager 1 and 2's Imaging Science System (ISS) \citep{Smith_VoyagerImagingExperiment_SSR_1977} i.e. two cameras (narrow and wide angle) per spacecraft. Images have a 1024x1024 pixel resolution for LORRI, 800x800 for SSI and 1000x1000 for ISS. This work has been done in the context of the preparation of ESA's JUICE mission \citep{Grasset_JUpiterICymoons_PaSS_2013} and NASA's Europa Clipper \citep{Phillips_EuropaClipperMission_ETAGU_2014} but the proposed strategy is general enough to be used for every past and future space exploration missions. It could have many applications such as vision-based navigation \citep{Jonniaux_AutonomousVisionBased_IAC_2016}, spectroscopic and photometric studies \citep{Belgacem_RegionalStudyEuropas_I_2019}. 

\begin{figure*}[ht]
\centerline{\includegraphics[width=16cm]{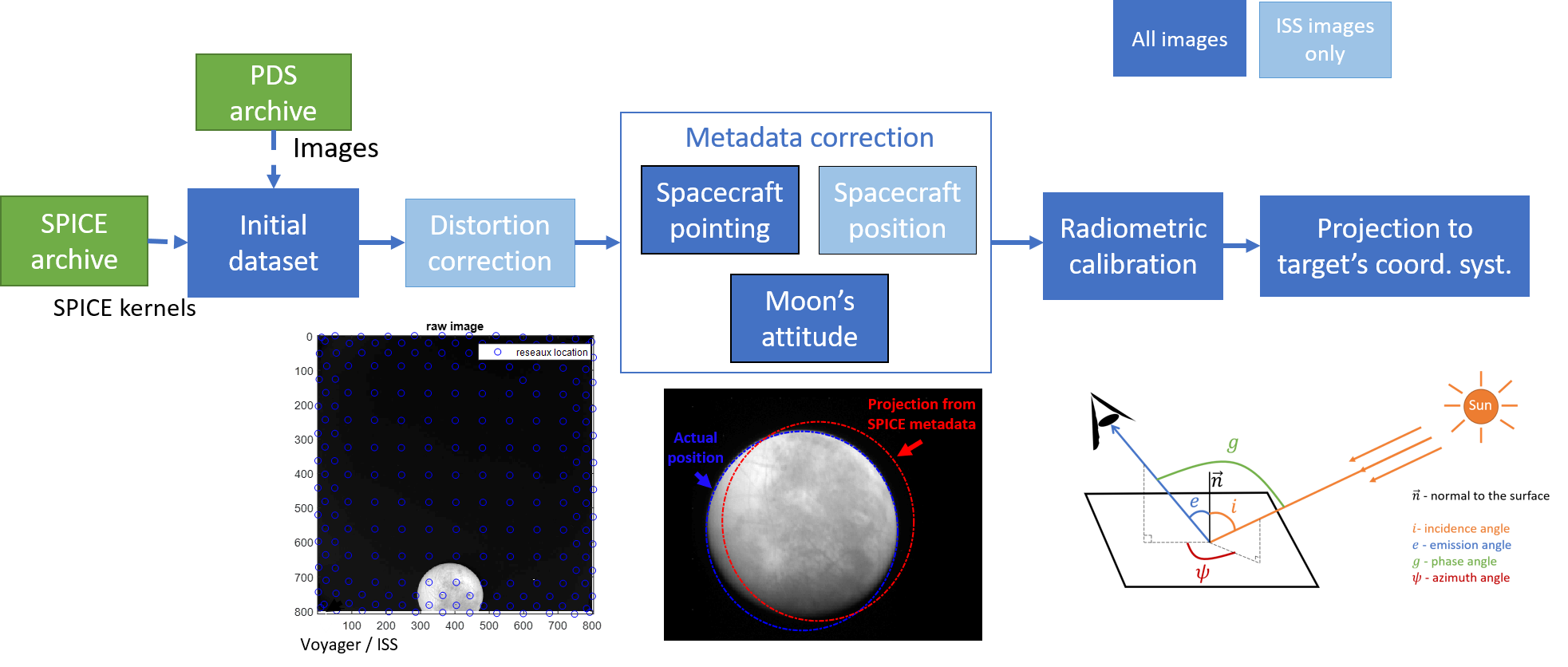}}
\caption{Visualization of entire pipeline to correct metadata and extract geometrical information from the images}
\label{fig:pipeline}
\end{figure*}

Fig. \ref{fig:pipeline} summarizes all the different steps of our approach. We propose several alternatives and choose the most reliable solution for each brick of the pipeline.

\section{Distortion}
Before any kind of metadata correction, it is necessary to address the optical distortion of the camera in the images. The most common type of distortion is radial \citep{Hartley_ComputationCameraMatrix_2003}. It is symmetric and can follow one of two patterns - "barrel", "pincushion" - or a combination of the two. Some cameras such as the ones embarked on Voyager can have a more complex behavior.

\subsection{LORRI}
The geometric distortion has been estimated to be less than $0.5$ pix across the entire field of view \citep{Cheng_LongRangeReconnaissance_SSR_2008}. We considered that there was no need for correction.

\subsection{Galileo SSI}
The distortion of the Galileo SSI instrument is well known and expected to be very stable. It is a negative radial distortion ("pincushion") and is maximal at the corners of the image with about $1.2$pix and increases from the center of the image as the cube of the distance \citep{Belton_GalileoSolidState_SSR_1992}. Therefore we can easily correct the images using this power law. For each point in the raw image noted $(x,y)$ or $(r, \theta)$ in central cylindrical coordinates, we can compute its undistorted position $(x_d, y_d)$ or $(r_d, \theta_d)$ using:

\begin{equation}
\begin{array}{ccc}
r_d = r + \delta_r & with & \delta_r = 1.2 \times \left(\frac{d}{d_{c}}\right)^3
\end{array}
\end{equation}

Where $d$ is distance from the center and $d_{c}$ is distance of any corner from center: $d_{c} = 400\sqrt{2}$

\begin{figure}[ht]
\centerline{\includegraphics[width=7cm]{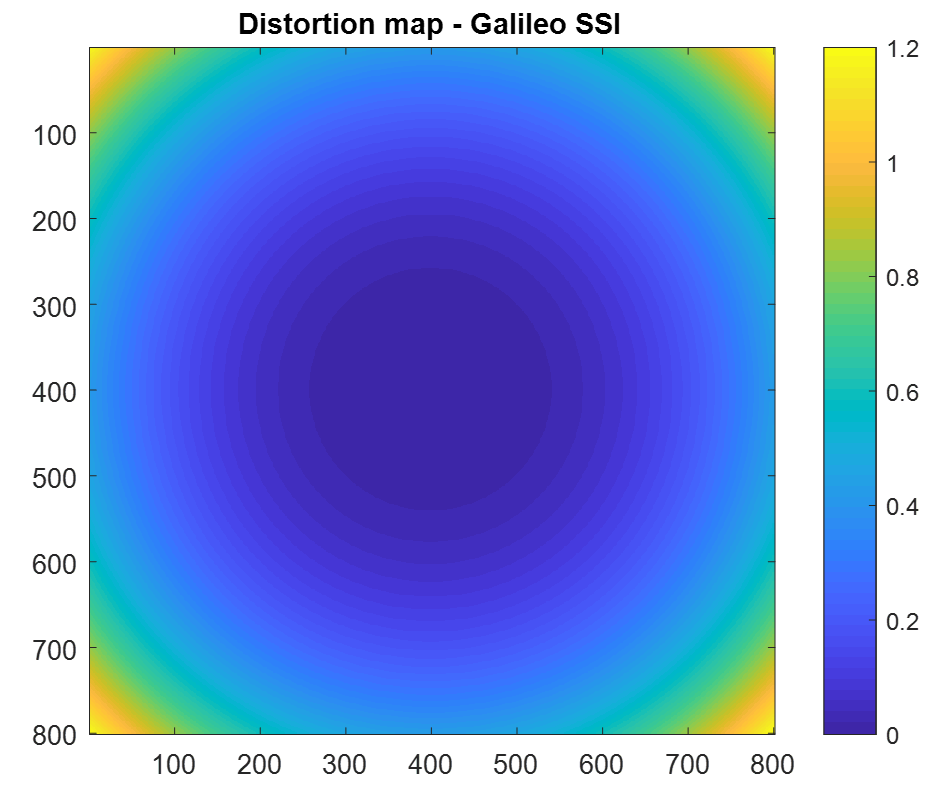}}
\caption{Variation of $\delta_r$ across the field of view of Galileo SSI}
\label{fig:distmap_galileo}
\end{figure}

Fig. \ref{fig:distmap_galileo} shows the distortion map of Galileo SSI, i.e. the value of $\delta_r$ across the field of view of the camera. A simple bilinear interpolation between the positions in the raw image and the corrected coordinates gives us the undistorted image.

We should note that another quadratic distortion correction has been proposed by \citealt{Oberst_VerticalControlPoint_JoGR_2004}.

\subsection{Voyager}

\begin{figure}[ht]
\centerline{\includegraphics[width=12cm]{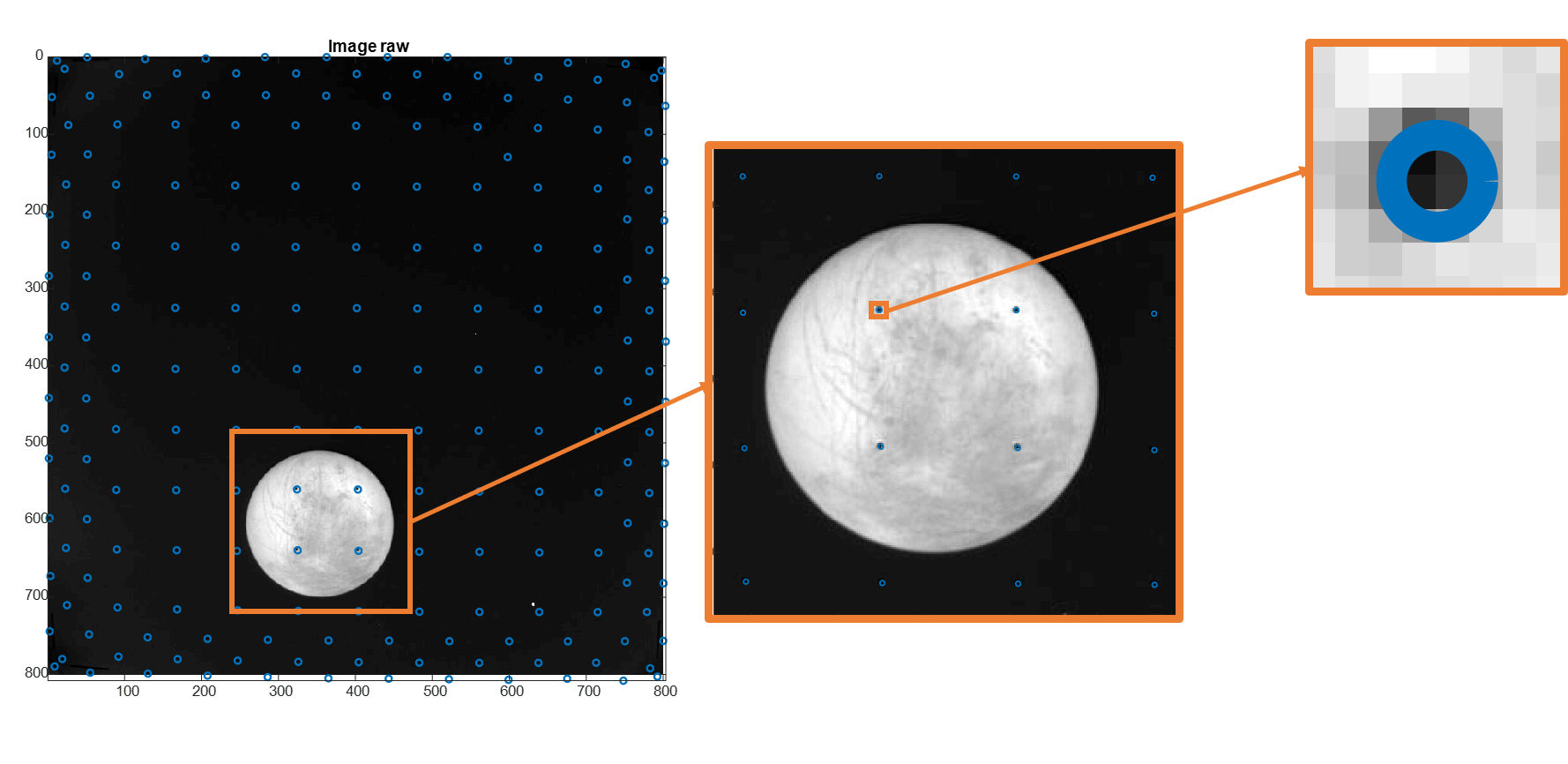}}
\caption{Raw Voyager image with reseau markings}
\label{fig:rawVG}
\end{figure}

Voyager is the first generation of space exploration from the 70's and the level of distortion in the images is much higher. It is a lot less stable than in other data sets and requires a specific processing. A grid of reseau markings was embedded within the optics of the cameras to record it. Fig. \ref{fig:rawVG} shows a raw Voyager 1 NAC image where we can see the markings. The real positions of the reseau were measured during calibration and are supposed to be fixed. On each image, the apparent position of the reseau can be measured. We can then correct for distortion using the correspondence. This work has been done by the NASA Planetary Data System Ring Node service and their corrected images are available online \citep{Showalter_PDS_Voyager_archive}. We have noticed that some residual distortion was still present so we propose here to improve the correction. We used the measured positions tabulated in the \textit{geoma} files provided by \citealt{Showalter_PDS_Voyager_archive} that were quite precise (see zoomed image in fig. \ref{fig:rawVG}). We propose the 3 following methods that we quantify by their residual RMSD in the prediction of the reseau points of image C1636902. 

\subsubsection{Method 1 - radial solution}

Let's consider $(x,y)$ the pixel coordinates in the raw image and $(x_d,y_d)$ the pixel coordinates in the undistorted image, $r=\sqrt{x^2+y^2}$ the radial coordinate. We are looking for the radial distortion function $f$ defined by \citep{Hartley_MultipleViewGeometry_book_2003}:

\begin{equation}
\label{eq:rad_func2}
r_d = rf(r_1)=r(1+k_1r+k_2r^2+k_3r^3+ \cdots)
\end{equation}

We use a least-squares approach to estimate the $k_i$ coefficients. Considering that each of the $n$ reseau point will contribute two equations ($x_d =  xf(r)$ and $y_d = yf(r)$) we have a system of $2n$ equations that we can write $A\mathbf{X} = \mathbf{b}$. The least squares solution is given by:

\begin{equation}
\mathbf{X}=(A^TA)^{-1}A^T\mathbf{b}
\end{equation}

We could not achieve satisfactory results with this approach. The best result we achieved using a radial distortion function was an average RMSD (Root Mean Square Deviation) of $17.3$ pix using a 9-degree polynomial in $r$ over 382 Voyager images. The distortion in the Voyager images cannot be described as radial and needs to be addressed in a more specific manner. The exhaustive list of images can be found in the supplementary material.

\subsubsection{Method 2 - general polynomial}

More general 2-variables-polynomial functions allow for a non radial symmetric distortion:

\begin{equation}
\label{eq:dist_pol}
\begin{array}{c}
x_d = k_{10} + k_{1a}x + k_{1b}y + k_{1c}xy + k_{1d}x^2 + k_{1e}y^2 
\\
y_d = k_{20} + k_{2a}x + k_{2b}y + k_{2c}xy + k_{2d}x^2 + k_{2e}y^2 
\end{array}
\end{equation}

We solve for the $k_i$ coefficients with a least-squares method and make several attempts with eq. \ref{eq:dist_pol} and higher degree polynomials. Results are improved and the best performance is achieved with a 6-degree polynomial - we obtain an average RMSD of $0.47$pix over the same 382 Voyager images. The exhaustive list of images can be found in the supplementary material.

It is worth noting that these systems are not very stable because the matrix A we define is close to being singular. This problem is more likely to happen the higher the degree of the polynomial.

\subsubsection{Method 3 - local bilinear transformations}
Another option is to work on a more local level. We start by dividing the reseau markings grid into triangles using the Delaunay algorithm. For each of them, we compute the exact bilinear transformation that would transform it into its undistorted form.  

For $T$, a triangle between three points $t_1=\begin{pmatrix}x_1 \\ y_1 \end{pmatrix}$ $t_2=\begin{pmatrix}x_2 \\ y_2 \end{pmatrix}$ and $t_3=\begin{pmatrix}x_3 \\ y_3 \end{pmatrix}$ whose undistorted equivalent is $T'$, a triangle between $t_1'=\begin{pmatrix}x'_1 \\ y'_1 \end{pmatrix}$, $t_2'=\begin{pmatrix}x'_2 \\ y'_2 \end{pmatrix}$ and $t_3'=\begin{pmatrix}x'_3 \\ y'_3 \end{pmatrix}$ we can write:

\begin{equation}
\label{eq:xyi}
\begin{array}{rcl}
x_i' = ax_i + by_i + c
\\
y_i' = dx_i + ey_i + f
\end{array}
\forall i\epsilon \{1,2,3\}
\end{equation}

If we form the vector $\mathbf{X_A}^T = \begin{pmatrix}
a & b & c & d & e & f
\end{pmatrix}$, we can rewrite eq. \ref{eq:xyi} as:

\begin{equation}
\label{eq:xyi_mat2}
\begin{array}{rcl}

\begin{pmatrix}
x_i & y_i & 1 & 0 & 0 & 0 \\
0 & 0 & 0 & x_i & y_i & 1
\end{pmatrix}
\begin{pmatrix}
a \\ b \\ c \\ d \\ e \\ f
\end{pmatrix}
=
\begin{pmatrix}
x'_i \\ y'_i 
\end{pmatrix}

\\
~
\\

A_x X_A = B 

\end{array}
\end{equation}

Each point of the triangle contributes two lines to the matrix $A_x$. To find the coefficients of the transformation matrix stored in the vector $\mathbf{X_A}$, we need to invert the system. The least squares solution is given by:

\begin{equation}
\label{eq:sol_triang}
\mathbf{X_A} = (A_x^{T}A_x)^{-1}A_x^{T}B
\end{equation}

The undistorted image is a 1000*1000 square. Each pixel can be attributed to a triangle and undergo the corresponding transformation to compute its position in the raw image. With a simple bilinear interpolation between the raw image and the projected grid of undistorted positions, we have the new undistorted image. The RMSD cannot be used to evaluate the precision since the the reseau point are perfectly matched by construction. Thus we cannot have a precise estimation of the accuracy of this correction method beyond the fact that it is under a pixel if, as we suppose, the distortion is well sampled by the reseau grid. 

\begin{figure}
\centerline{\includegraphics[width=16cm]{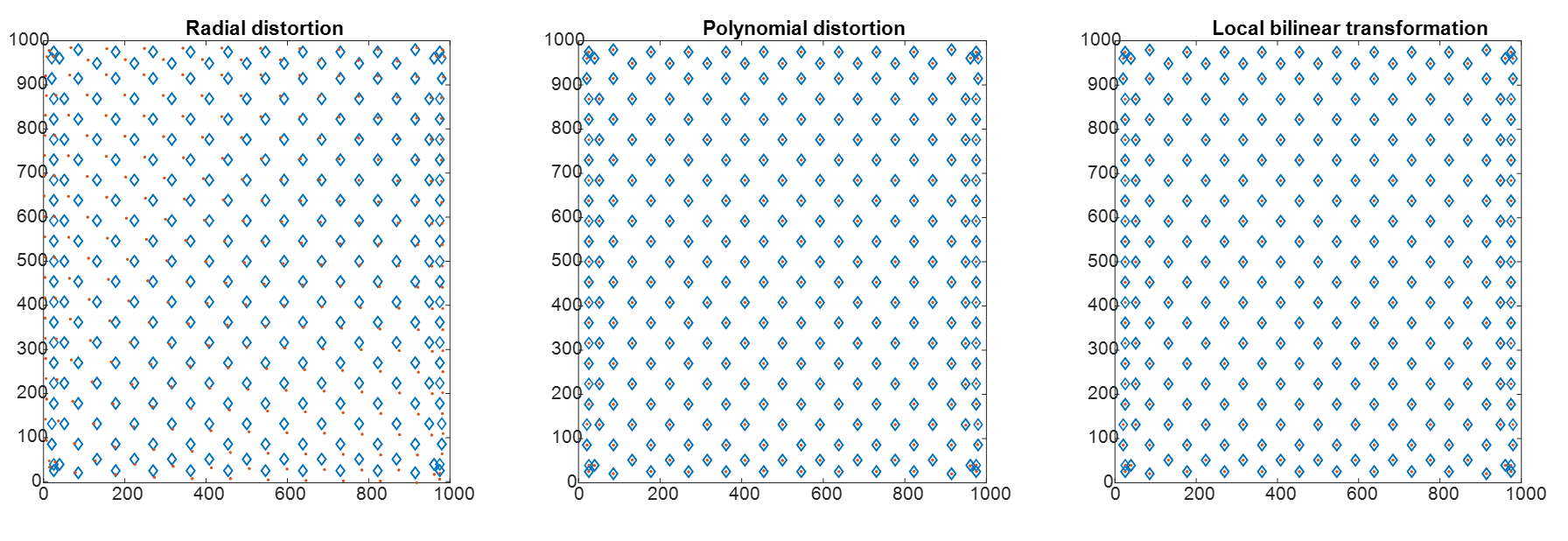}}
\caption{Illustration of the matching of the reseau markings computed location in the 1000*1000 grid using the three methods we present here (magenta dots) to their known location (blue diamonds) for Voyager 1 NAC image C1636858.}
\label{fig:reseau_loc}
\end{figure}

We choose to apply the local bilinear transformation method since it is the one that ensures perfect reconstruction for the full reseau markings grid. Fig. \ref{fig:reseau_loc} illustrates the matching of the reseau markings computed location in the 1000*1000 grid using the three methods we present here to their known location. 

\section{Camera pose}

In this section, we address the pointing error as an error on the camera pose (spacecraft position and orientation). We should mention that other potential sources of errors could come from uncertainties in the definition and alignment of the reference frame of the instrument, or its boresight. Table \ref{tab:all_offsets} will show that there is a significant spread in our pointing corrections which would not be consistent with a systematic equal bias on all images for a given camera. 

\subsection{Spacecraft Pointing}
\begin{figure}[ht]
\centerline{\includegraphics[width=7cm]{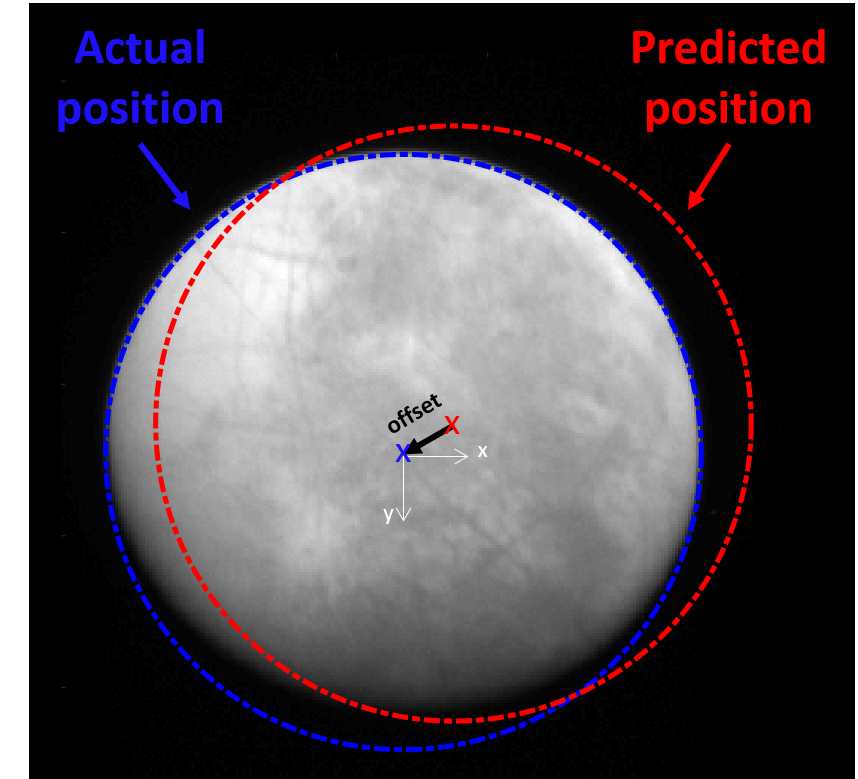}}
\caption{Illustration of remaining pointing error on a LORRI image}
\label{fig:simuvreel_pointing}
\end{figure}

To estimate the pointing error of the instrument, we can use the metadata available via the C-Kernels to project the shape of the target into the field of view of the camera. This gives us the predicted position of Europa in the image space in red in fig. \ref{fig:simuvreel_pointing} and we can see that it is a few pixels away from its actual position (in blue).  We have to keep in mind that even a slight error in pointing could result in an offset of tens of pixels in the image. Such errors are not surprising given typical star tracker accuracy (a few arcseconds \citep{Fialho_TheoreticalLimitsStar_S_2019, Eisenman_NewGenerationAutonomous_SPIE_1997} which for LORRI is equivalent to a few pixels in each direction). In this case, the center position of the moon is offseted by -16.5 pixels on the x-axis and -9.4 pixels on the y-axis. More illustrations of the pointing corrections are available for all data sets in the supplementary material. 

\subsubsection{2D analysis - measure the offset}

We are using SurRender, an image renderer developed by Airbus Defence $\&$ Space \citep{Brochard_ScientificImageRendering_IAC_2018}, to simulate images using the metadata from the SPICE kernels and compare them to the real images (fig. \ref{fig:simuvreel_pointing}). In these conditions, the simplest approach would be to consider a normalized cross-correlation to compute the translation in the field of view between the two images using:

\begin{equation}
\label{eq:cross_corr}
\rho = \frac{\sum_{x,y} \left[ f_{r}(x,y)-\overline{f_{r}} \right]
\left[ f_{s}(x, y)-\overline{f_{s}} \right]} 
{\sqrt{\sum_{x,y} \left[ f_{r}(x,y)-\overline{f_{r}} \right]^2 
\sum_{x,y} \left[ f_{s}(x, y)-\overline{f_{s}} \right] ^2}} 
\end{equation}

With :
\begin{itemize}
\item $f_{r}$: real image, $\overline{f_{r}}$: mean real image
\item $f_{s}$: simulated image, $\overline{f_{s}}$: mean simulated image
\end{itemize}

However, this solution at best gives pixel-scale errors.

\begin{figure}[H]
\centerline{\includegraphics[width=10cm]{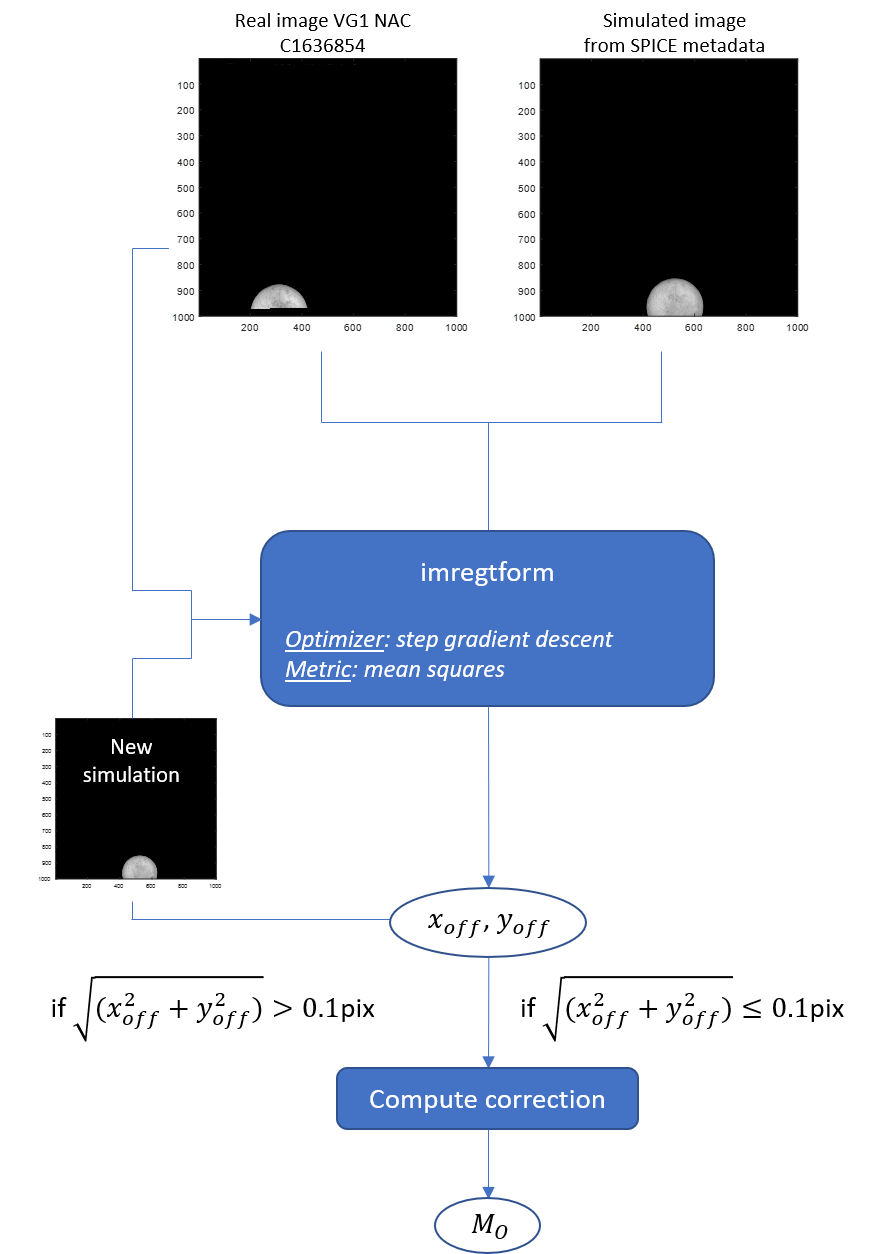}}
\caption{Algorithm used to estimate the translation in the image frame}
\label{fig:imreg_algo}
\end{figure}

To have a better estimate of the pointing error, we finally chose to use an optimization-based method using intensity-based registration. The function we used performs a regular step gradient descent and uses a mean squares metric to compare the two images at each step (imregtform function, \citep{Mathworks_MatlabImageProcessing_2018}). We use this function in a loop to ensure that we determine the offset between real image and simulation down to a 1/10th of a pixel i.e. we repeat the process and update the simulation with the corrected camera orientation until the offset computed between simulation and real image is under 0.1 pixel as described in fig. \ref{fig:imreg_algo}.

\begin{figure}[H]
\centerline{\includegraphics[width=14cm]{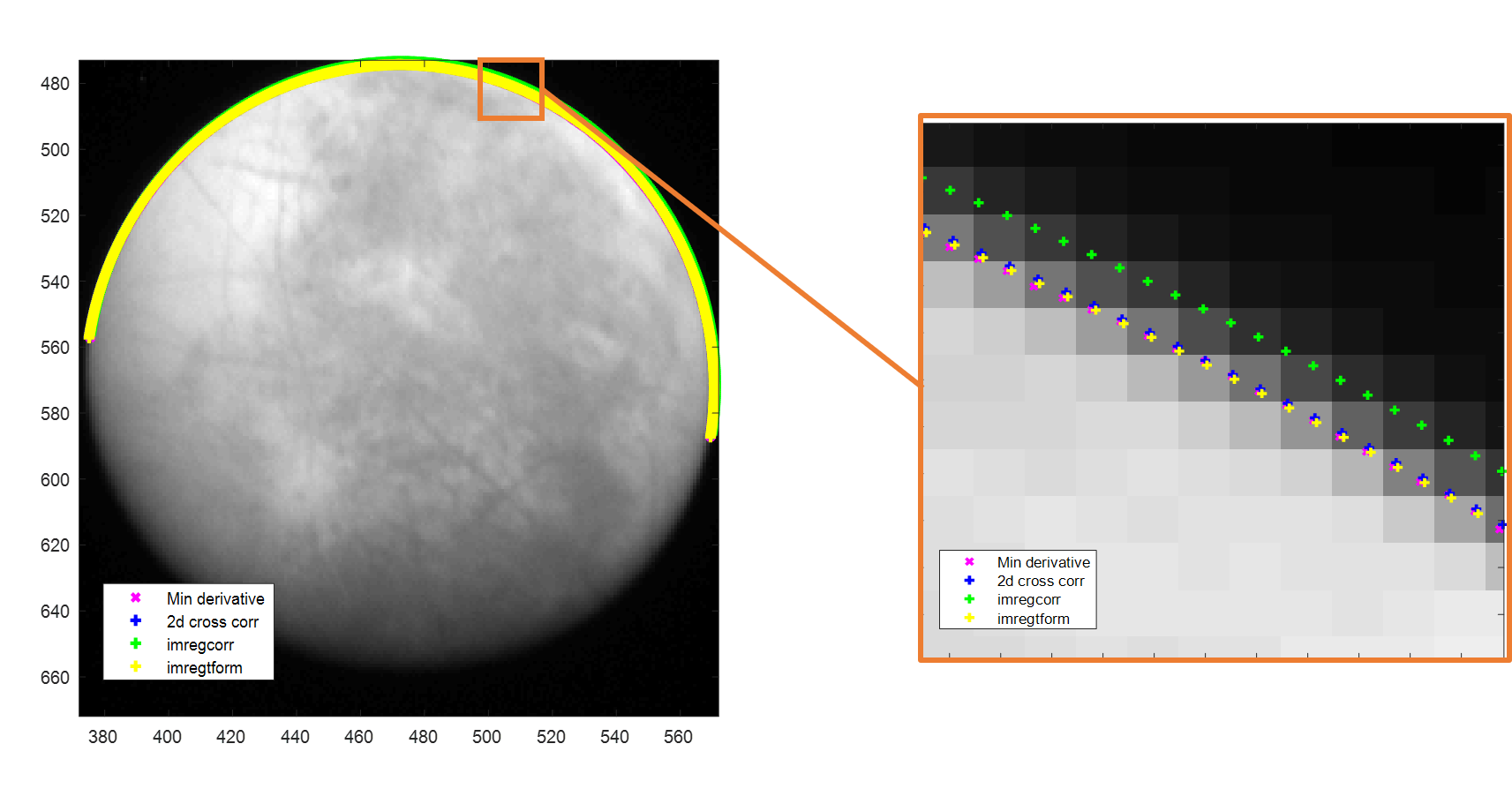}}
\caption{Illustration of different limb corrections on New Horizons' LORRI image LOR\_0034849319\_0X630\_SCI\_1}
\label{fig:limbcorr}
\end{figure}

This is the most precise method we have explored and above all the most robust regardless of the data set. This step can be replaced by any other method of image registration \citep{Luo_FastNormalizedCross_IToUFaFC_2010, Ma_RobustFeatureMatching_IToGaRS_2015, Reddy_FFTBasedTechnique_IToIP_1996} without impacting the rest of the pipeline as long as an offset is computed. We illustrate that on fig. \ref{fig:limbcorr} by comparing the corrected limb with different methods: 2D cross correlation (in blue), \textsc{MATLAB} function imregcorr (in green), \textsc{MATLAB} function imregtform (in yellow) and a limb detection method (in magenta).

We should also note that the methods we considered were all tested on entirely synthetic situations. As an example, let’s take a simulated image of Europa. We have introduced an offset of 20.32 pixels in the x-direction and an offset of -30.46 pixels in the y-direction, represented by the initial limb in red. We retrieved these offsets using three different methods and the results are summarized in table \ref{tab:simu_imreg}.


\begin{table}[h]
	\centering
	
	\begin{tabular}{|c|c|c|c|c|}
		\hline
		~ & Theory & 2D cross-correlation & imregcorr & imregtform \\
		\hline
		xoffset (pix) & 20.32 & 20.0 & 20.3 & 22.33 \\
		\hline
		yoffset (pix) & -30.46 & -30.4 & -30.4 & -30.48 \\
		\hline
	\end{tabular}
	
	\caption{Results after trying to retrieve a synthetic offset using three different methods}
	\label{tab:simu_imreg}
\end{table}

\begin{figure}[H]
\centerline{\includegraphics[width=18cm]{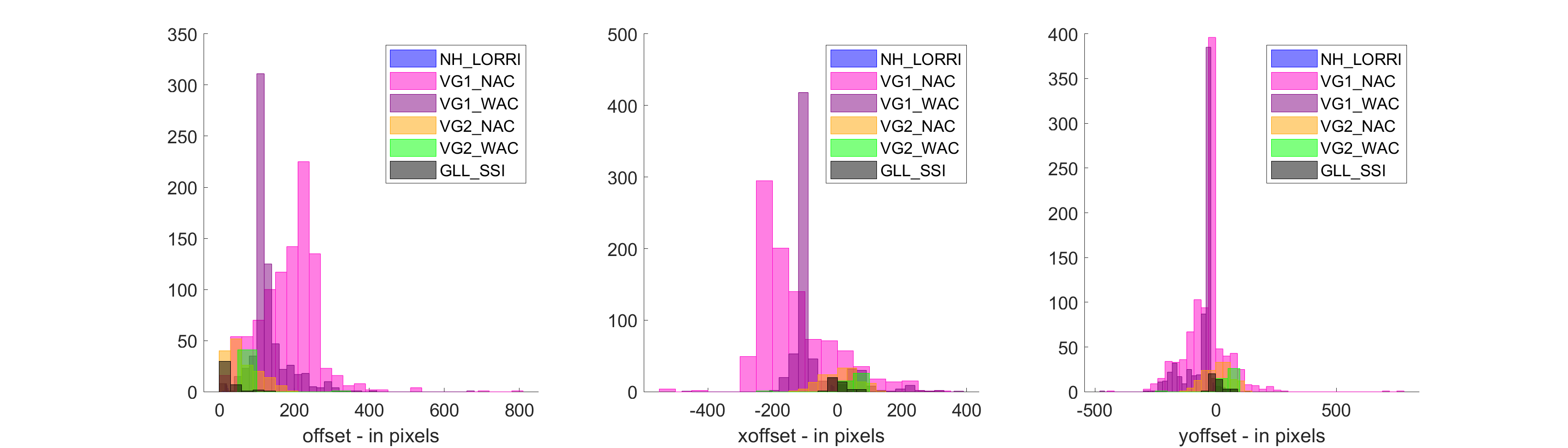}}
\caption{Distribution of pointing errors in pixels for the different cameras}
\label{fig:histVoyOffset}
\end{figure}

Table \ref{tab:all_offsets} summarizes the values of the offset we computed for the entirety of the data sets of Europa images. We have highlighted in bold the mean values in pixel. We can see that the metadata (extrinsic parameters) errors are much more substantial in the Voyager images. Their distribution is more detailed in fig. \ref{fig:histVoyOffset}. We can see that beyond the fact that the values are significant, the distributions are shifted in the negative values.

\begin{table}[h]
	\centering
	
	\begin{tabular}{|c|c|c|c|c|c|c|c|c|c|}
		\hline
		\multicolumn{2}{|c}{} &  \multicolumn{2}{|c}{Voyager 1} & \multicolumn{2}{|c|}{Voyager 2} & New Horizons & Galileo \\
		\cline{3-8}
		\multicolumn{2}{|c|}{} & NAC & WAC & NAC & WAC & LORRI & SSI \\
		\hline
		\multirow{3}{*}{offset} & \textbf{mean} & \textbf{186.1} & \textbf{131.1} & \textbf{63.5} & \textbf{81.2} & \textbf{18.1} & \textbf{22.4} \\
		\cline{2-8}
								 & min  & 0.0 & 7.2 & 1.2 & 52.5 & 11.0 & $1e^{-5}$ \\
		\cline{2-8}
								 & max  & 796.4 & 675.9 & 191.1 & 340.5 & 23.4 &  120.9 \\
		\hline
		\multirow{3}{*}{xoffset} & mean & 160.9 & 104.8 & 44.9 & 57.4 & 15.0 & 15.9 \\
		\cline{2-8}
								 & min  & 0.0 & 2.0 & 0.9 & 37.1 & 3.8 & $1e^{-5}$ \\
		\cline{2-8}
								 & max  & 537.2 & 478.2 & 135.1 & 240.8 & 22.6 & 85.5 \\
		
		\hline      
		\multirow{3}{*}{yoffset} & mean & 64.6 & 63.6 & 44.9 & 57.4 & 9.4 & 15.77 \\
		\cline{2-8}
								 & min  & 0.0 & 0.0 & 0.9 & 37.1 & 6.0 & $1e^{-5}$ \\
		\cline{2-8}
								 & max  & 777.0 & 447.7 & 135.1 & 240.8 & 13.6 & 85.5 \\
		\hline
	\end{tabular}
	
	\caption{Summary of all the offsets computed on images of Europa taken with the different cameras}
	\label{tab:all_offsets}
\end{table}

\subsubsection{Computing the 3D transform}
Once the offset computed in the image space, we need to find the associated 3D transform (rotation) to actually correct for spacecraft pointing. For that, we need to consider the expected boresight of the instrument corrected by the computed offset in the image:

\begin{equation}
\overline{\mathbf{b}} = K^{-1}\left[ \begin{pmatrix} x_{off} \\ y_{off} \end{pmatrix} + \begin{matrix} c_x \\ c_y \end{matrix} \right]
\end{equation}

Where:
\begin{itemize}
\item $K$: the intrinsic matrix of the camera \citep{Hartley_CameraModels_2003} describing the field of view and focal of the camera
\item $x_{off}, y_{off}$: offsets along the x and y-axis
\item $c_x, c_y$: pixel coordinates of the camera center
\end{itemize}

\begin{figure}[ht]
\centerline{\includegraphics[width=10cm]{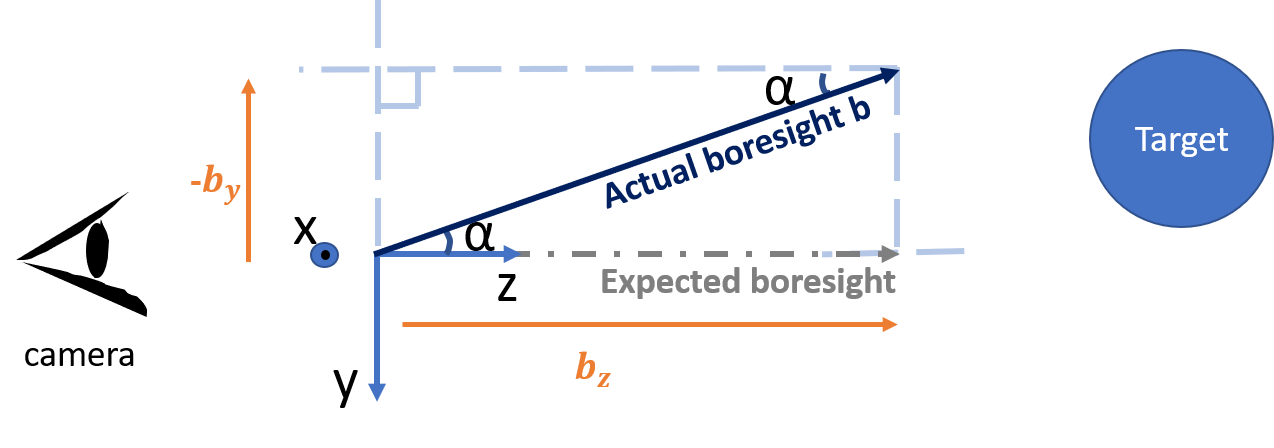}}
\caption{Visualization of boresight in the yz plane}
\label{fig:bsight_yz}
\end{figure}

$\overline{\mathbf{b}} $ is the normalized vector (in the camera frame) representing the actual boresight of the camera. To derive the correcting Euler angles , we can use fig. \ref{fig:bsight_yz} showing the yz plane. Around the x-axis: $\alpha = arctan\left(\frac{-b_y}{b_z}\right)$. A similar approach leads to deriving the angle around the y-axis $\beta = arctan\left(\frac{b_x}{b_z}\right)$. 

The rotation matrix $M_{o}$ associated to the Euler angles $[\alpha, \beta, 0]$ is the correction factor we apply to the camera orientation to correct the pointing. Please note that we do not estimate the rotation around the boresight which will be addressed in section \ref{sec:target_attitude}.

\subsection{Distance}

\begin{figure}[ht]
\centerline{\includegraphics[width=12cm]{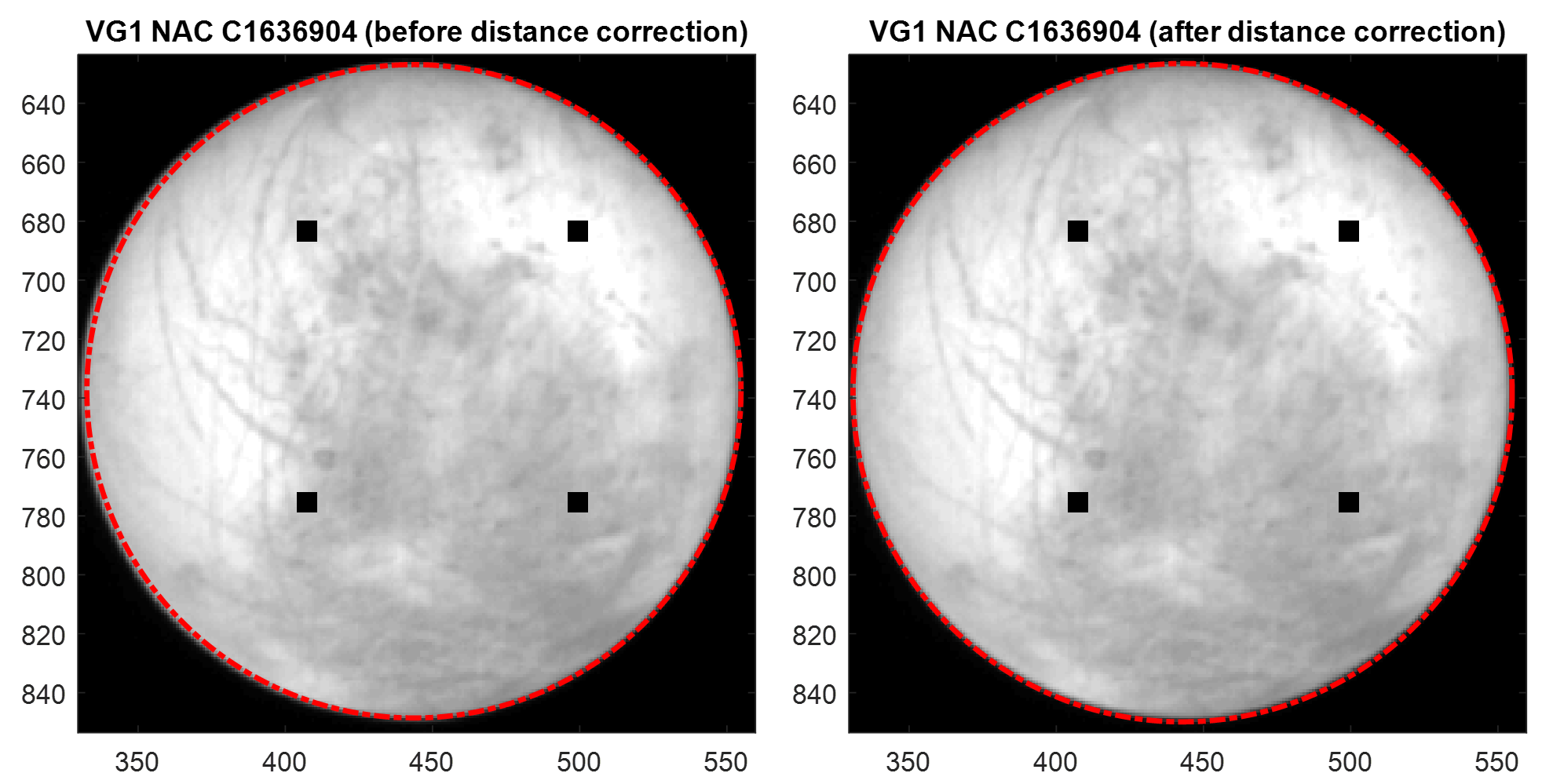}}
\caption{VG1 NAC image with Europa out of limb bounds, especially on the left of the moon (left) and the result after distance correction (right)}
\label{fig:lune_deborde}
\end{figure}

Fig. \ref{fig:lune_deborde} shows an example of a Voyager observation, after correction of spacecraft pointing, where we can see that the moon is actually expanding beyond the limb in red.

This can be the result of either an underestimated field of view, or distance. We found that this effect is dependent on the image, thus it is more likely that a correction of distance is needed. We continue using comparisons to simulated images, the changing parameter being, this time, the distance between camera and target. 

To evaluate the match between real and simulated images we chose the Structural SIMilarity index (SSIM) which robustness to noise was noted by \citep{Loza_StructuralSimilarityBased_MVaA_2007} who used it as a tracking technique in videos. The SSIM of two images $a$ and $b$ is defined by the combination of three terms - luminance ($l(a,b)$), contrast ($c(a,b)$) and structure ($s(a,b)$):

\begin{equation}
ssim(a,b) = l(a,b)c(a,b)s(a,b)
= \left[\frac{2\mu_a\mu_b}{\mu_a^2 + \mu_b^2}\right] \left[\frac{2\sigma_a\sigma_b}{\sigma_a^2 + \sigma_b^2}\right] \left[\frac{\sigma_{ab}}{\sigma_a\sigma_b}\right]
\end{equation}

Where $\mu$ is the sample mean, $\sigma$ is the standard deviation and $\sigma_{ab}$ is the sample covariance. We use the SSIM index to define our cost function. To minimize the cost function (or maximize the similarity index), we try doing a simple gradient descent but the algorithm was thrown off by local minima. We decided to resort to a less optimized but safer method: computing the cost function for a set of distance values between 99$\%$ and 101$\%$ of the predicted distance and picked the distance for which we had the best index a posteriori.

\begin{figure}[H]
\centerline{\includegraphics[width=12cm]{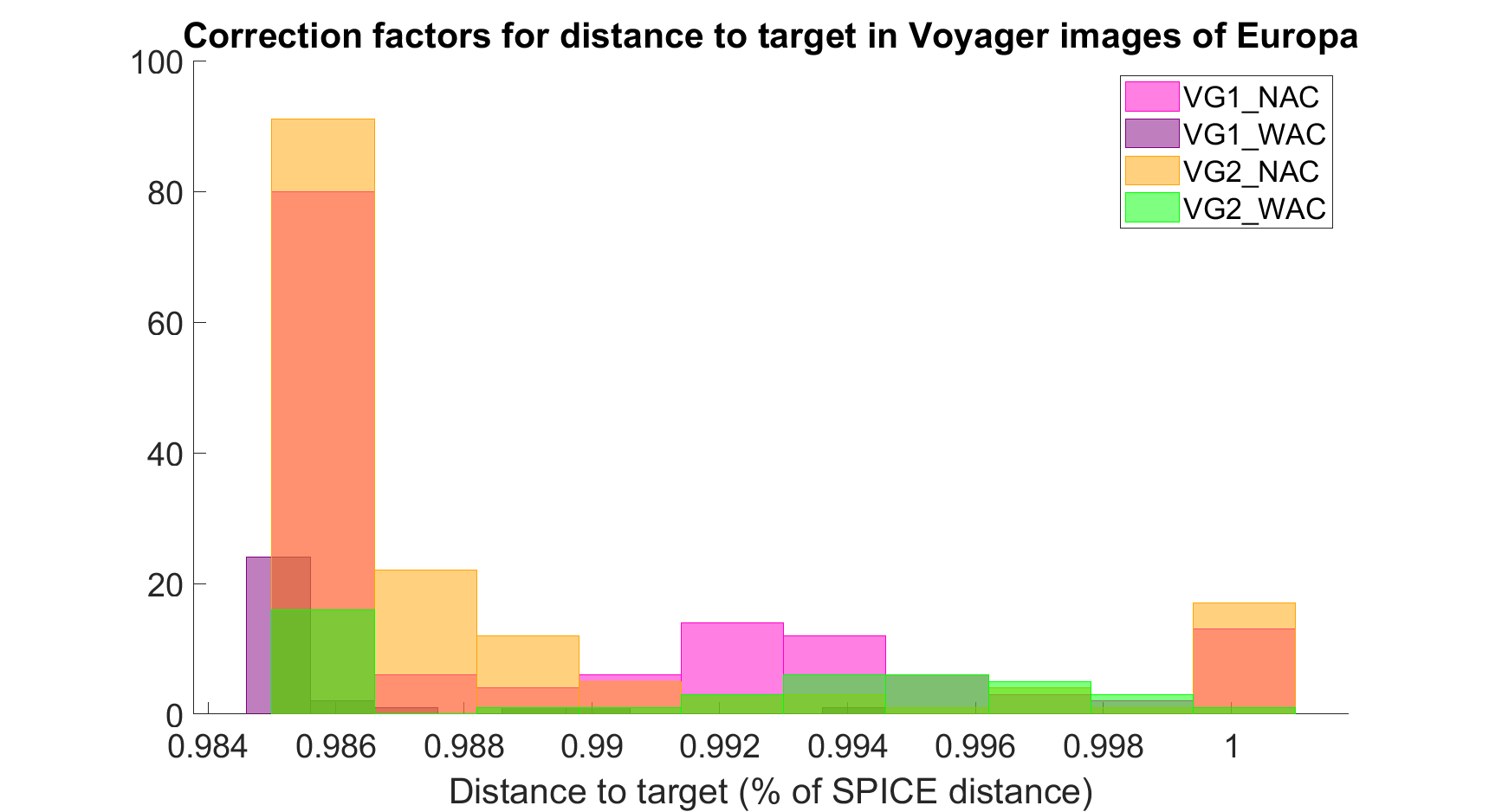}}
\caption{Distribution of correction factors for the distance to target in Voyager images of Europa}
\label{fig:hist_dist_corr}
\end{figure}

Fig. \ref{fig:hist_dist_corr} shows the distribution of the computed correction factors over the totality of the Voyager data set for Europa. We can note that the distance has always been overestimated and that Europa is actually always closer than expected from the SPICE kernels. 

\section{Validation}

\subsection{Distortion}

\begin{figure}[H]
\centerline{\includegraphics[width=15cm]{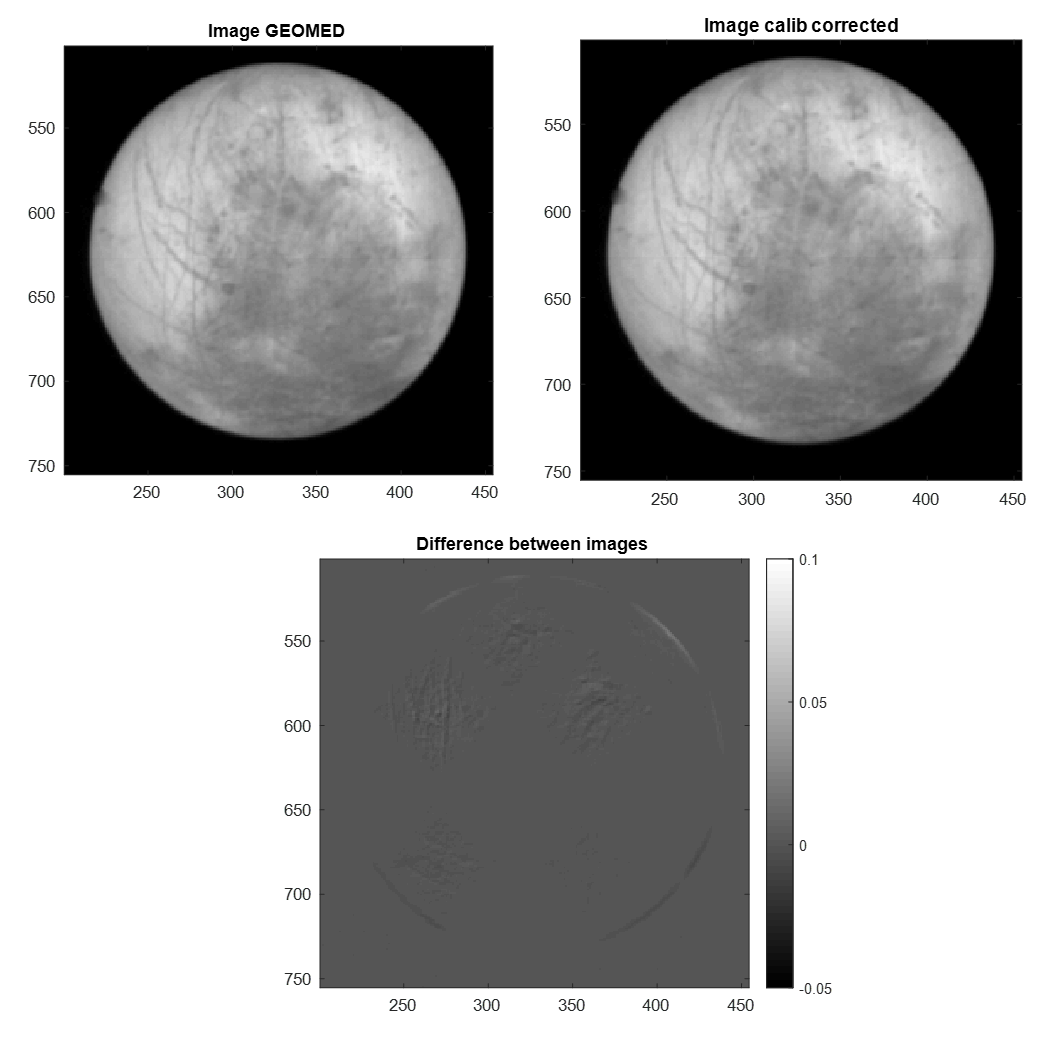}}
\caption{Illustration of the differences between our undistorted images and the GEOMED images made available by the PDS Ring Node \citep{Showalter_PDS_Voyager_archive} with Voyager 1 NAC image C1636902.}
\label{fig:comp_distortion}
\end{figure}

\begin{figure}[H]
\centerline{\includegraphics[width=15cm]{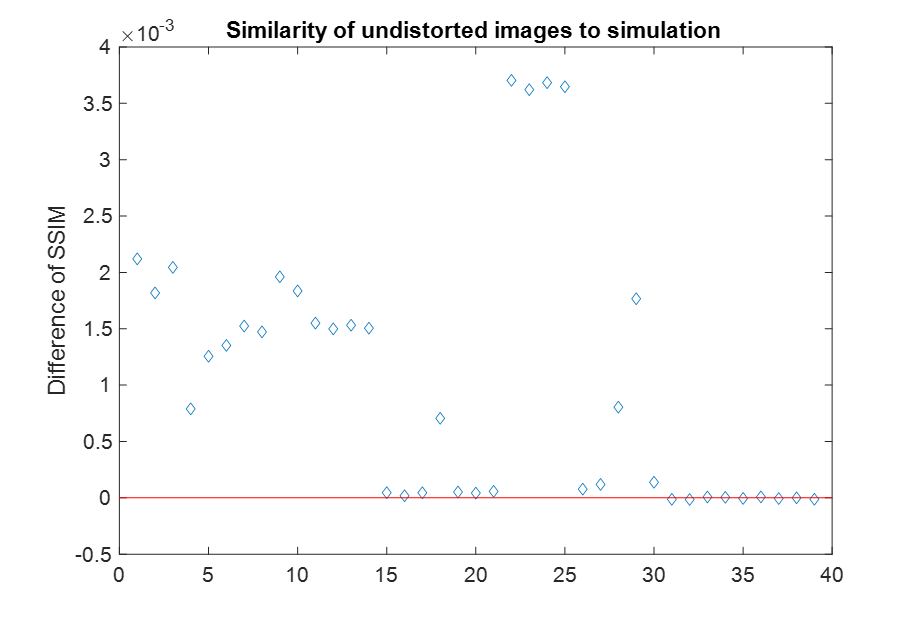}}
\caption{Values of the difference between the structural similarity index (ssim) comparing our undistorted images and a simulation and the ssim value comparing the GEOMED images and the same simulation. The absolute values of the index are around 0.99. The images represented here are the 39 Voyager images used in \citealt{Belgacem_RegionalStudyEuropas_I_2019} and listed in the supplementary.}
\label{fig:ssim_distortion}
\end{figure}

We compared our undistorted images using a local bilinear transformation (see section 2.3.3) to the GEOMED images made available in the Ring Node archive \citep{Showalter_PDS_Voyager_archive}. Fig. \ref{fig:comp_distortion} and \ref{fig:ssim_distortion} illustrate the differences we have noted.

We should note that our correction is only slightly better than the GEOMED images (images corrected for distortion by VICAR software). One way of looking at it is comparing the GEOMED image and our undistorted image to a simulation. When doing that, we find that our correction gives more similar images (quantified with the structural similarity index) to the simulation than the GEOMED. However, we are looking at differences in the index of the order of 1e-3 and ssim values around 0.99. 

\subsection{Camera pose}

For a complete validation of the camera pose, we look closely at the new predicted limb, after the different corrections. Although we are strictly looking at the limb here, we do not only validate the correction of the spacecraft orientation: if the distortion is not corrected or if the camera position is still wrong, it will also show as a poorly corrected limb. 

\begin{figure}[H]
\centerline{\includegraphics[width=8cm]{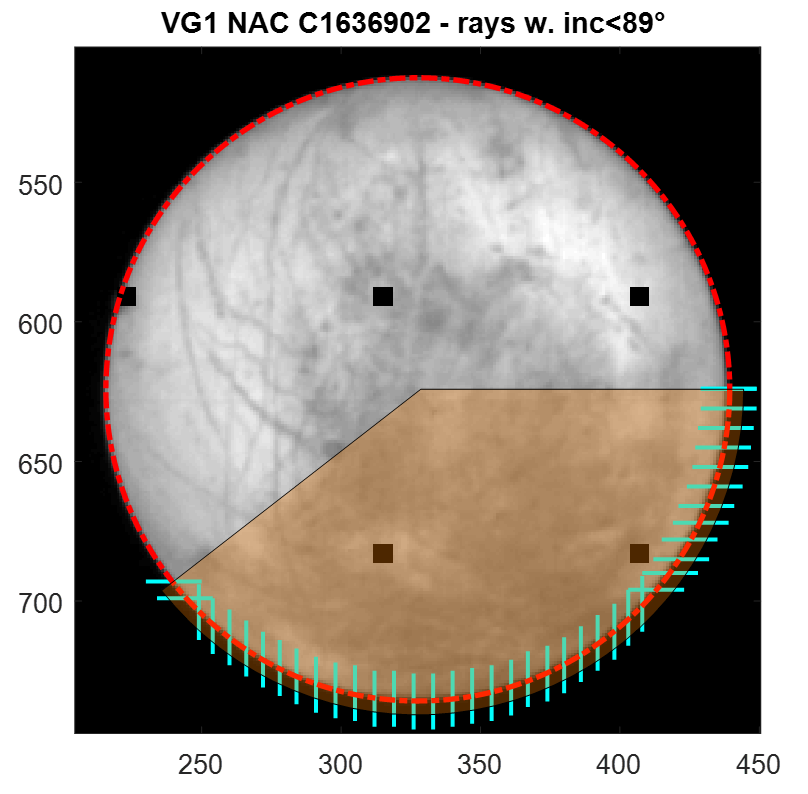}}
\caption{Example of a Voyager 1 NAC image with segments drawn on the illuminated part of the limb}
\label{fig:ill_seg}
\end{figure}

For each point on the limb of the target, we can trace two segments in the image frame - one vertical and one horizontal. We only keep the one less tangent to the limb. The relevant information is in the fully illuminated part of the limb (highlighted in orange in fig. \ref{fig:ill_seg}). 

\begin{figure}[H]
\centerline{\includegraphics[width=16cm]{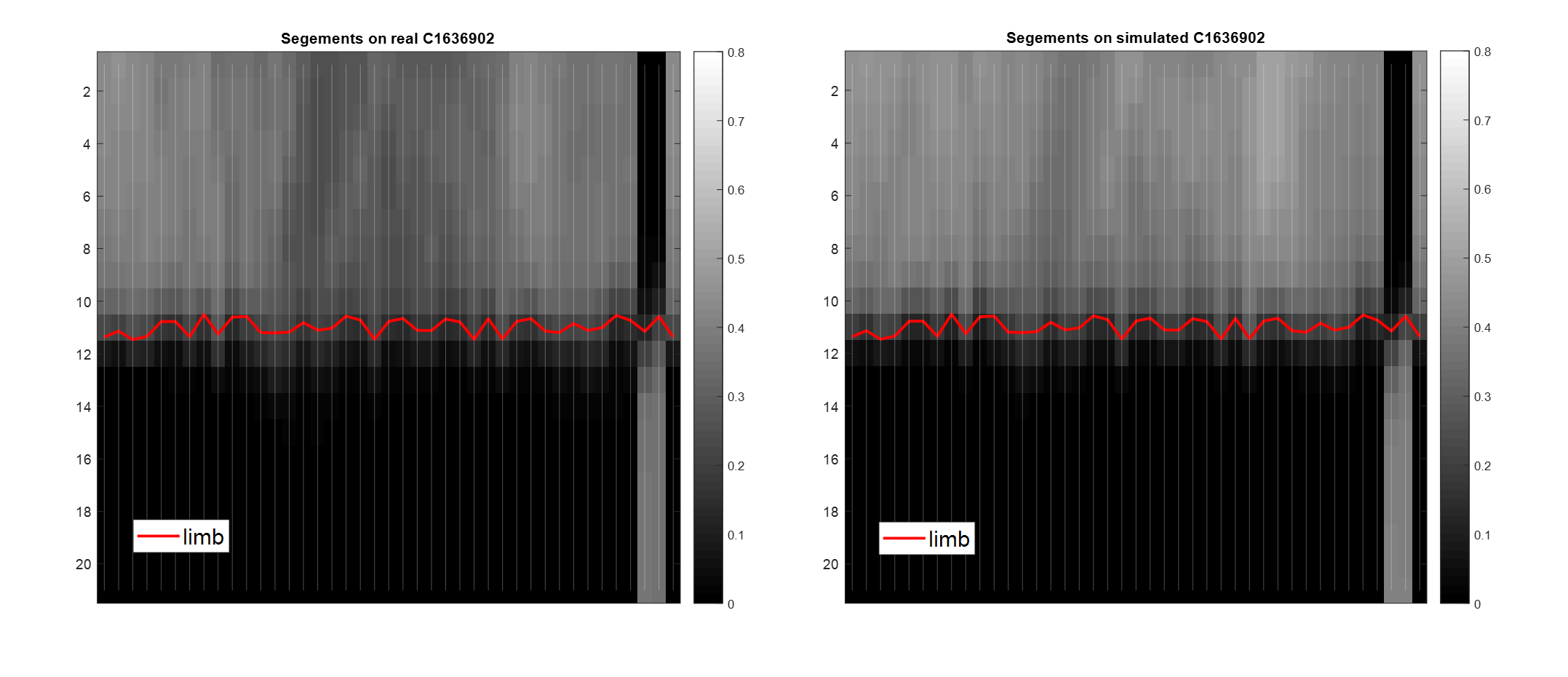}}
\caption{Comparison of limb segment between real image and simulation. In red is the newly corrected limb after metadata correction.}
\label{fig:seg_img}
\end{figure}

\begin{figure}[H]
\centerline{\includegraphics[width=16cm]{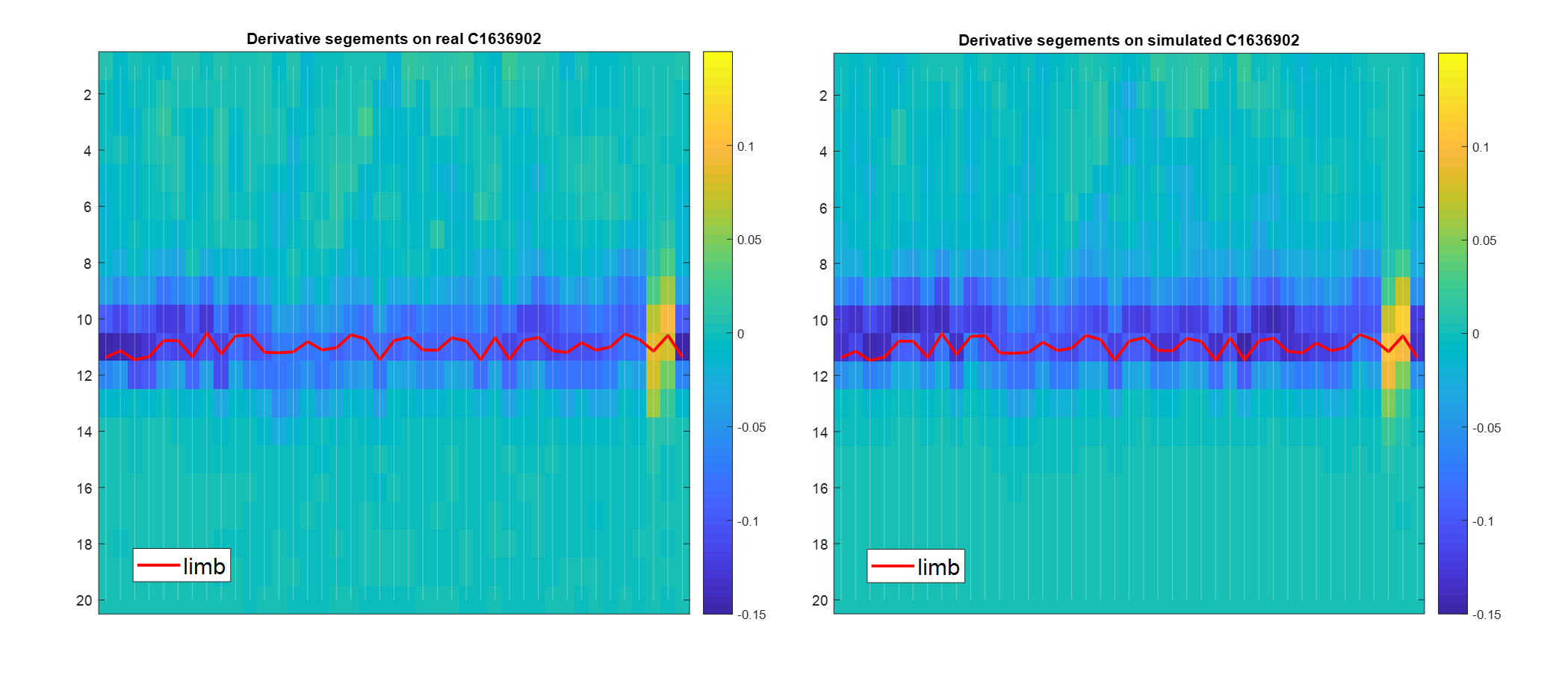}}
\caption{Comparison of derivatives of limb segment between real image and simulation. In red is the newly corrected limb after metadata correction.}
\label{fig:dseg_img}
\end{figure}

Each of these segments gives a piece of information about the position of the limb. If we visualize each of them individually we can see quite precisely how well the new predicted limb fits the target on the image. The limb is supposed to be at the extremum of the derivative of the segment: it is the strongest change from the illuminated target to the blackness of the background sky. Fig. \ref{fig:seg_img} shows the illuminated segments displayed on fig. \ref{fig:ill_seg} compared to their simulated equivalent. Fig. \ref{fig:dseg_img} shows the derivatives.

\begin{figure}[H]
\centerline{\includegraphics[width=15cm]{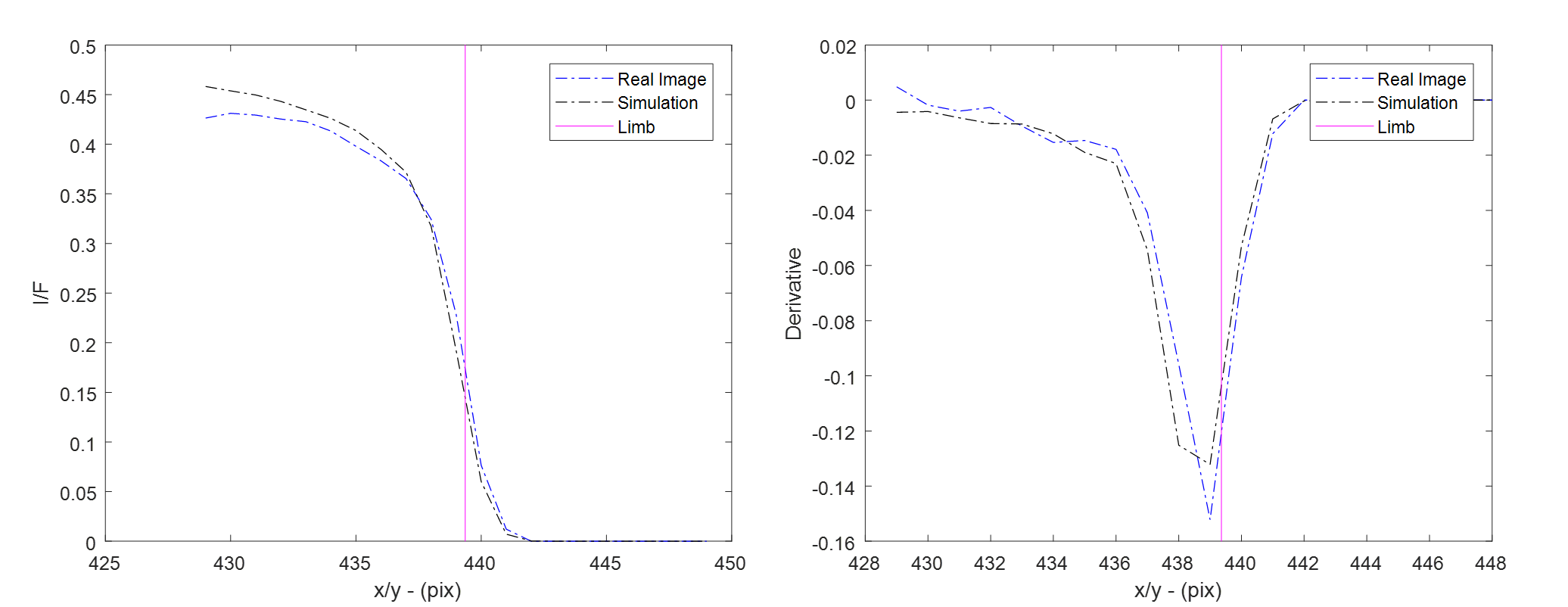}}
\caption{Example of a segment and its derivative after complete correction of camera pose}
\label{fig:seg_match}
\end{figure}

\begin{figure}[H]
\centerline{\includegraphics[width=15cm]{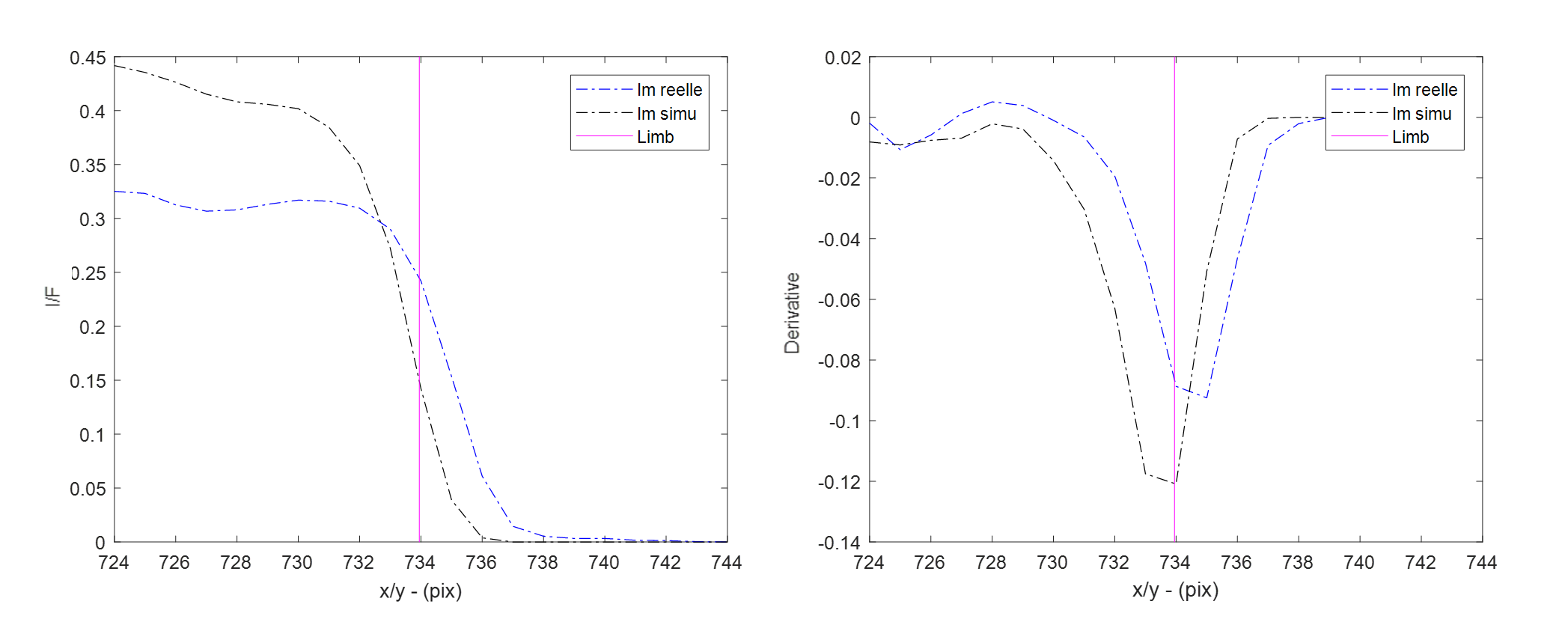}}
\caption{Example of a segment and its derivative with a remaining error in the camera pose. In red is the newly corrected limb after metadata correction.}
\label{fig:seg_nmatch}
\end{figure}

The form of the derivative is not trivially described. A first order approximation would be a Gaussian but a few tests showed quickly that it was not enough. That is why we chose to compare each segment to its equivalent in the simulation. If the new predicted limb fits perfectly, both derivatives should have the same extremum (fig. \ref{fig:seg_match}). However, if the camera pose is still off, this will show as a shift between simulation and reality both in the segment itself and its derivative (fig. \ref{fig:seg_nmatch}).

\section{Target's attitude}
\label{sec:target_attitude}
After correcting the camera pose (attitude + position with respect to Europa), some differences remain between the images and their simulations. We perform an optical flow measurement to interpret these differences. At least some of them can be explained by an imprecision of the moons' attitude with respect to the camera. There is indeed a global movement that can be corrected by a slight rotation of the target - Europa here. The rotation around the boresight of the camera - not corrected so far - can also contribute to that effect. 

\subsection{2D analysis - optical flow}
\begin{figure}[h]
\centerline{\includegraphics[width=15cm]{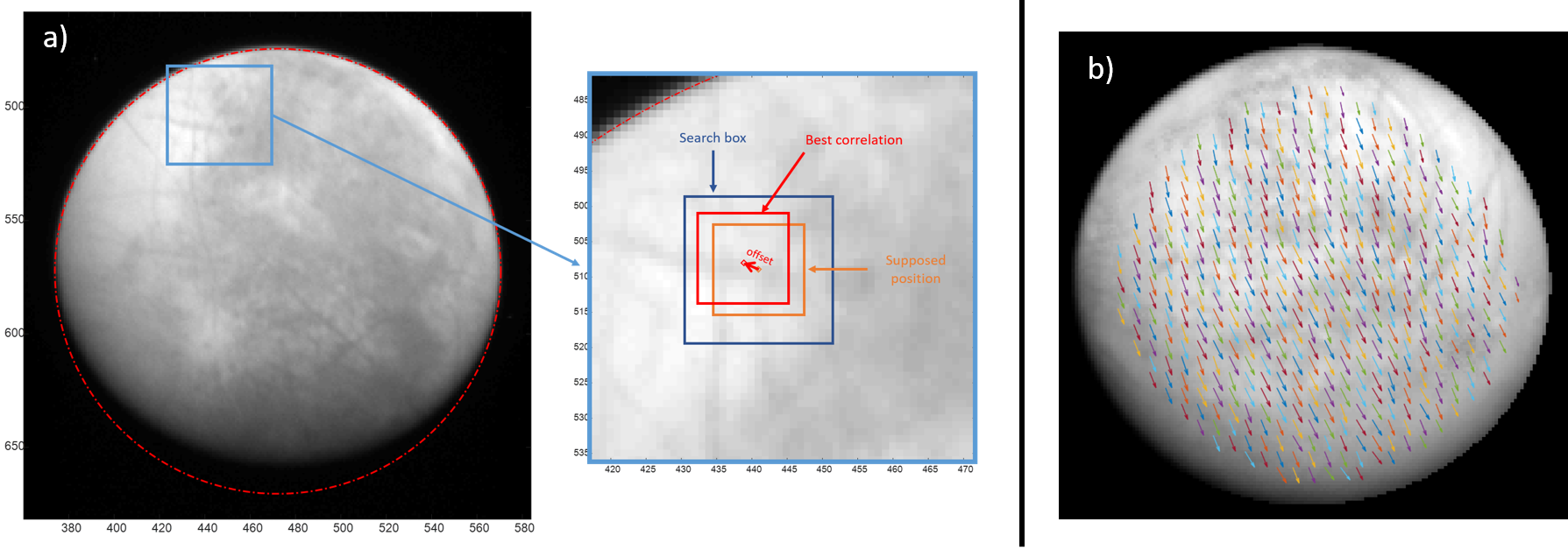}}
\caption{a) Definition of ROI and search box for a pixel b) Example of optical flow resulting from a simulated rotation of Europa}
\label{fig:optical_flow}
\end{figure}

An optical flow describes the apparent motion of an object in an image. We choose a local approach: for each pixel we define a region of interest consisting of a 13-pixel square box in the simulated image and search for the best normalized cross-correlation in a 21-pixel square box in the real image (fig. \ref{fig:optical_flow}a) using equation \ref{eq:normxcorr2}. 

\begin{equation}
\label{eq:normxcorr2}
\rho_{u,v} = \frac{\sum_{x,y} \left[ f(x,y)-\overline{f}_{u,v} \right]
\left[ t(x-u, y-v)-\overline{t} \right]} 
{\sqrt{\sum_{x,y} \left[ f(x,y)-\overline{f}_{u,v} \right]^2 
\sum_{x,y} \left[ t(x-u, y-v)-\overline{t} \right] ^2}} 
\end{equation}

Where:
\begin{itemize}
\item $t$ is the ROI in the simulated image, $\overline{t}$ is the mean of the ROI in the simulated image
\item $f$ is the search box in the real image, $\overline{f}_{u,v}$ is the mean of the search box in the equivalent ROI in the real image
\end{itemize}

Thus, for each pixel, we have a displacement vector pointing to the best local correlation between the simulated and real images (fig. \ref{fig:optical_flow}b). At the image scale, we obtain a global pattern of displacement indicating in which direction the moon has to be rotated in the simulated image to better match the real one.

More details on the optimization of this process and the choice of the reference map that was used to generate the images are available in \citealt{Belgacem_PhotometricStudyJupiters_PhDThesis_2019} (chapter 4, sections 4.3 and 4.4).

\subsection{Computing the 3D solid rotation - Kabsch algorithm}
We need to compute the correcting rotation associated with the movement pattern. From the optical flow, we have two sets of matching 2D points. We first project these points to obtain two sets of 3D points. We choose to implement Kabsch algorithm \citep{Kabsch_DiscussionSolutionBest_ACSA_1978} to compute the rotation that minimizes the RMSD between the two sets of points. Let's represent both sets of $n$ points by matrices $P$ and $Q$

\begin{equation}
\begin{array}{cc}
P=\begin{pmatrix}
x_1 & y_1 & z_1 \\
x_2 & y_2 & z_2 \\
\vdots & \vdots & \vdots \\
x_n & y_n & z_n 
\end{pmatrix}_{set\#1}
&
Q=\begin{pmatrix}
x_1 & y_1 & z_1 \\
x_2 & y_2 & z_2 \\
\vdots & \vdots & \vdots \\
x_n & y_n & z_n 
\end{pmatrix}_{set\#2}
\end{array}
\end{equation}

Then, we compute the covariance matrix $C = P^TQ$. The matrix $C$ is not necessarily inversible, we thus need to use the single value decomposition. Let $U$, $\Sigma$ and $V$ be the matrices of this decomposition such as $C = U\Sigma V^T$. Finally, the rotation matrix that best matches the two sets of points $P$ and $Q$ is given by:

\begin{equation}
R=V \begin{pmatrix}
1&0&0\\0&1&0\\0&0&det(VU^T)
\end{pmatrix}
U^T
\end{equation}

To ensure that R is expressed in a direct right-handed coordinate system, we need $det(VU^T)>0$. If it is not the case, we have to invert the sign of the last column of matrix $V$ before calculating $R$. This rotation matrix is the correction factor to apply to the target's attitude.

We have to note that the choice of the texture is decisive in this approach. For instance, in the case of Europa, we used the color map \citep{Jonsson_MappingEuropa_2015} to produce the simulated images. If the map is erroneous, every measurement made in comparison to the simulations will be erroneous as well. We have identified very few patches that seem to be badly registered with respect to the rest of the map but do not affect the measurements overall.

\section{Projections: camera to scene}
After correction of all the metadata, we can safely project each pixel of the images onto the target to compute the corresponding coordinates (latitude and longitude) and observation geometry (incidence, emission and phase angles).

We are modeling Europa as an ellipsoid. Each point $X=\begin{pmatrix} x & y & z \end{pmatrix}^T$ on the surface  verifies the equation: 
\begin{equation}
\label{eq:ellipsoid}
(X-V)^TA(X-V)=1
\end{equation}

Where:
\begin{itemize}
\item $S$: spacecraft position
\item $X$: point on the ellipsoid
\item $V$: center of the ellipsoid
\item $A$: positive definite matrix parametrising the quadric
\end{itemize}

$A$ is a parametrisation matrix in any arbitrary orientation of the quadric. In the principal axes of the ellipsoid, it can be simplified to 
$A = \begin{pmatrix}
\frac{1}{r_e} & 0 & 0 \\
0 & \frac{1}{r_e} & 0 \\
0 & 0 & \frac{1}{r_p} 
\end{pmatrix}$
where $r_e$ is the equatorial radius of the ellipsoid and $r_p$, the polar radius.

We chose to express every coordinate in the J2000 frame which means that we will use:

\begin{equation}
A = R_T^{-1} \begin{pmatrix}
\frac{1}{r_e} & 0 & 0 \\
0 & \frac{1}{r_e} & 0 \\
0 & 0 & \frac{1}{r_p} 
\end{pmatrix} R_T
\end{equation}

Where $R_T$ is a matrix that transforms any coordinates in the target's fixed frame to J2000 - i.e. the target's attitude depending on time.

Each pixel of the detector has a line of sight - a 3-D vector. In order to project all the pixels onto the moon, we target to compute the intersection of these lines of sight with the ellipsoid modeling the planetary body. This is equivalent to solving equation \ref{eq:ellipsoid} after replacing $X$ by:

\begin{equation}
X=S+kL
\end{equation}

Where $k \epsilon \mathbb{R}$ is the distance from the pixel to the target and
$L$ is a $3 \times N$ matrix of unitary vectors, each being the line of sight of a pixel on the detector.

We obtain:
\begin{equation}
(S+kL-V)^T A (S+kL-V) = 1
\end{equation}

We have a second degree equation to solve for k that - once developed - can be written:

\begin{equation}
(L^TAL)k^2 + (2S^TAL-2L^TAV)k + (V^TAV - 2S^TAV + S^TAS - 1) = 0
\end{equation}

We can compute the determinant by:

\begin{equation}
\Delta = (2S^TAL-2L^TAV)^2 - 4(L^TAL)(V^TAV - 2S^TAV + S^TAS - 1)
\end{equation}

Three cases can arise:
\begin{itemize}
\item $\Delta<0$: no solution, the line of sight doesn't intersect the ellipsoid, the pixel doesn't see the target
\item $\Delta=0$: one solution, the pixel intersects the target on its exact edge
\item $\Delta>0$: two solutions, the line of sight intersects the ellipsoid twice, on the spacecraft-facing side and on the other side of the target along the same axis. In this case, we keep the closest point (spacecraft-facing face) which means the lowest $k$.
\end{itemize}

Solving the equation gives us the exhaustive collection of pixels in a position to "see" the moon. We still need to eliminate the pixels seeing the night side of the moon. To do so, we compute the geometry of observation at each intersection and eliminate all pixels seeing an area where the incidence angle is greater than 90\degree.

This approach shows a clear advantage compared to existing functions in SPICE - SINCPT, ILLUMIN, RECLAT - that are not vectorized. A vectorized projection of the entire 1024$\times$1024 pixel grid of a New Horizons' LORRI image took 0.45 seconds compared to a limiting 1 minute and 33 seconds using the SINCPT function in a loop. We should mention that the PDS Ring-Moon systems node has also developed a Python toolbox - OOPS - that simplifies the use of these SPICE functions \citep{Showalter_DevelopmentsGeometricMetadata_IaDA_2018}.

\section{Conclusion}
We have developed a complete pipeline to process images and convert them into usable and precise science products for a variety of applications. As an example of application, we have used these tools in a regional photometric study of Europa \citep{Belgacem_RegionalStudyEuropas_I_2019} for which an accurate projection of the individual pixels in the images was crucial to obtain the right coordinates and geometry of observation. We successfully ran the pipeline in its entirety on the full pertinent collection of 57 images taken with New Horizons' LORRI and Voyager's ISS. An exhaustive list of the images used in \citealt{Belgacem_RegionalStudyEuropas_I_2019} is available in the supplementary material. As a future work, we will compute and make our corrected metadata available for the Europa images at this link: \url{https://github.com/InesBlgcm/ImageProcessing}. We will also reach out to SPICE experts to generate C-smithed kernels for the relevant data set. Our vectorized solution for projecting pixels onto an ellipsoid target will also be very useful to estimate the geometry efficiently. 

We have to note that our approach is dependent on a reliable image renderer and most of all a reliable texture for the target, especially for correcting the target's attitude. Without these resources a less precise pointing correction would still be possible using a projected ellipsoid in the field of view in place of a more thorough simulated image. 

Another major hypothesis is to consider the ephemeris of the planetary bodies involved to be perfectly known. An improved approach would also correct for planetary ephemeris. This could be achieved with a more general use of a software such as CAVIAR \citep{Cooper_CaviarSoftwarePackage_AA_2018} that is for now dedicated to correcting CASSINI's ISS images. After a first correction based on background stars, our image processing approach would enable an improved knowledge of the ephemeris of the planetary bodies in the field of view. 

Although we have carried out this work with images of Europa, this approach should be easily adaptable on any other target. We validate here the pipeline on images from six different cameras, demonstrating its versatility. We also could imagine carrying out a similar approach for small bodies as long as a precise shape model is available to simulate our images.

\section*{Acknowledgment}
This work is supported by Airbus Defence $\&$ Space, Toulouse (France) as well as the "IDI 2016" project funded by the IDEX Paris-Saclay, ANR-11-IDEX-0003-02 and the ``Institut National des Sciences de l'Univers'' (INSU), the Centre National de la Recherche Scientifique (CNRS) and Centre National d'Etudes Spatiales (CNES) through the Programme National de Planétologie. We would also like to thank Mark Showalter for his insight on the Voyager data set as well as two anonymous reviewers for their valuable comments and feedback.

\bibliographystyle{elsarticle-harv}
\bibliography{PSS_bibtex}

\end{document}